\documentclass[12pt,preprint]{aastex}
\usepackage{epsfig}
\begin{document}
\voffset-1cm
\newcommand{\gsim}{\hbox{\rlap{$^>$}$_\sim$}}
\newcommand{\lsim}{\hbox{\rlap{$^<$}$_\sim$}}

\title{X-Ray Flares In GRB090812 - Case Study}

\author{Shlomo Dado\altaffilmark{1} and Arnon Dar\altaffilmark{2}}

\altaffiltext{1}{dado@phep3.technion.ac.il\\
Physics Department, Technion, Haifa 32000,Israel}
\altaffiltext{2}{arnon@physics.technion.ac.il\\
Physics Department, Technion, Haifa 32000, Israel}

\begin{abstract}

The master formulae of the CB model of long durations gamma ray bursts 
(GRBs) which reproduce very well the light curves and spectral evolution 
of their prompt emission pulses and their smooth afterglows, also 
reproduce very well the lightcurves and spectral evolution of their X-ray 
and optical flares. Here we demonstrate that for GRB090812 

\end{abstract}

\section{Introduction}

In more than 50\% of the gamma ray bursts (GRBs) observed with the Swift 
X-ray telescope (XRT), flares were observed at the end of the prompt 
emission and/or the early AG phase (see, e.g., Burrows et al.~2005;  
Burrows et al.~2007; Falcone et al.~2007). In some cases X-ray flares were 
observed also at very late times, of the order of several days after the 
prompt emission.  Flares in GRBs were studied phenomenologically by 
various observer groups (see, e.g., Burrows et al.~2005;  Burrows et 
al.~2007; Kocevski \& Butler~2007; Butler \& Kocevski~2007; Falcone et 
al.~2007;  Chincarini et al.~2007a,b,~2008a,b, and references therein). 
Modifications of previously suggested models and new theoretical models 
were proposed and discussed by several authors but none of the proposed 
models was shown to actually derive the observed spectral and temporal 
properties of either early-time flares or late-time flares from underlying 
physical assumptions except for the cannonball model of GRBs (see, e.g., 
Dado \& Dar, hereafter DD, 2009b and references therein).

Flares are a natural consequence of the cannonball (CB) model of GRBs, 
which was motivated by a GRB-microquasar analogy (e.g., Dar \& De 
R\'ujula 2004 and references therein, Dado, Dar \& De R\'ujula, hereafter 
DDD, 2002; 2009). In the CB model, {\it long-duration} GRBs and their AGs 
are produced by bipolar jets of highly relativistic plasmoids of ordinary 
matter ejected in accretion episodes on the newly formed compact stellar 
object (Shaviv \& Dar 1995; Dar~1998) in core-collapse supernova (SN) 
explosions (Dar et al.~1992; Dar \& Plaga 1999). It is hypothesized that 
an accretion disk or a torus is produced around the newly formed compact 
object, either by stellar material originally close to the surface of the 
imploding core and left behind by the explosion-generating outgoing shock, 
or by more distant stellar matter falling back after its passage (De 
R\'ujula 1987). As observed in microquasars, each time part of the 
accretion disk falls abruptly onto the compact object, two jets of 
cannonballs (CBs) made of {\it ordinary-matter plasma} are emitted 
with large bulk-motion Lorentz factors in opposite directions along the 
rotation axis wherefrom matter has already fallen back onto the compact 
object due to lack of rotational support. The prompt $\gamma$-ray and 
X-ray emission is dominated by inverse Compton scattering (ICS) of photons 
of the SN glory - scattered and/or emitted light by the SN filling the 
cavity produced by the pre-supernova wind/ejecta which was blown from the 
progenitor 
star long before the GRB. The CBs' electrons Compton up-scatter the glory 
photons into a narrow conical beam of $\gamma$ rays along the CBs' 
direction of motion. An X-ray `flare' coincident in time with a prompt 
$\gamma$-ray pulse is simply its low-energy part. The flares ending the 
prompt emission and during the early time afterglow is the same: ICS of 
glory photons by the electrons of CBs ejected in late accretion episodes 
of fall-back matter on the newly formed central object (e.g., Dar 2006). 
The early time X-ray flares without an accompanying detectable 
$\gamma$-ray emission are usually IC flares (ICFs) produced by CBs with 
relatively smaller Lorentz factors: As the accretion material is consumed, 
the `engine' has a few progressively-weakening dying pangs. Like the 
lightcurves of the prompt GRB pulses, the lightcurves of ICFs exhibit a 
rapid softening during their fast decline phase (see, e.g. Evans et al. 
2007,2009). Often, the fast decay of an ICF is taken over by 
synchrotron radiation (SR) from 
the CB's encounter with the wind enclosing the glory light before 
the take over by the plateau/shallow-decay phase of the afterglow (AG).

The initial expansion of the CBs and the slowing-down of the leading ones 
by the circumburst matter merge most of them during the afterglow phase 
into a single leading CB (Dar \& De R\'ujula 2004, DDD2009). The
prompt emission beam of gamma rays ionizes the matter in front 
of the CB. The ions continuously impinging on the CB with a relative 
Lorentz factor $\gamma(t)$, where $\gamma(t)$ is the bulk motion Lorentz 
factor of the CB, generate within it an equipartition turbulent magnetic 
field. The intercepted electrons are isotropized and Fermi accelerated by 
these fields and emit isotropic synchrotron radiation in the CB's rest 
frame, which is Doppler boosted and beamed relativistically into a narrow 
cone with a typical opening angle $\!\sim\!1/\gamma(t)$. Late 
 synchrotron radiation flares (SRFs) are 
produced mainly when the CBs encounter winds or density bumps along their 
path first from the progenitor star and later in the interstellar medium 
(ISM). The lightcurve of these flares depends on the unknown density 
profile of the encountered wind/density bump which cannot be predicted 
a-priori. But, both the early-time and the late-time SRFs have a typical 
SR spectrum and a weak spectral evolution which are quite different from 
those of the accretion induced ICFs and can be used to identify their 
origin -- late ejection episodes from the central engine or 
encounters with density bumps.

\section{The master formula for ICS flares}

Let  $t$ denote the time in the observer frame
after the beginning of a flare ($t\!=\!T\!-\!T_i$ where $T$ is the 
time after trigger and $T_i$ is its value  at the beginning 
of the flare. The light-curve of a flare, produced
by the electrons in a CB by ICS of
thermal bremsstrahlung photons filling the cavity formed by a wind
blown by the progenitor star long before the
GRB, is generally well approximated by (DDD2009 and references therein):
\begin{equation}
E\, {d^2N_\gamma\over dt\,dE}(E,t)\approx
A\, {t^2/\Delta t^2  \over(1+t^2/\Delta t^2)^2}\,
e^{-E/ E_p(t)}\,
\propto  e^{-E /E_p(0)}\, F(E\,t^2),
\label{ICSlc}
\end{equation}
where $A$ is a constant which depends on the CB's baryon number,
Lorentz and Doppler factors, and on the density
of the glory light and on the redshift and distance of the GRB,
and $E_p(t)$, the peak energy of $E^2\, d^2N_\gamma/ dE\, dt$
at time $t$ is given roughly by:
\begin{equation}
E_p(t)\approx  E_p(0)\, {t_p^2 \over t^2+t_p^2}\,,
\label{PeakE}
\end{equation}
with $t_p$ being the time (after the beginning of the flare) when the ICS
photon count-rate reaches peak value.
For $E\!\ll\!E_p$, it satisfies
$E_p(t_p)\!=\!E_p$ where $E_p$ is the
peak energy of the time-integrated spectrum of the flare.
Thus, in the CB model, each ICF in the GRB lightcurve
is described by four parameters, $A$,
$\Delta t(E),$  $E_p(0)$ and $T_i$, the
beginning time of the pulse when $t$ is taken to be 0.

The late-time decay of the energy flux of the prompt emission pulses
and ICFs in  an
energy band $[E1,E2]$, which follows from Eq.~(\ref{ICSlc}), is given
approximately by,
\begin{equation}
\int_{E1}^{E2} E\, {d^2N_\gamma\over dt\,dE}(E,t)\, dE\approx
A\, {E_p(t)\,\Delta t^2\over t^2}\,
[e^{-E1/E_p(t)}-e^{-E2/ E_p(t)}].
\label{InICSlc}
\end{equation}
Thus, for the Swift XRT lightcurves where $E1\!=\!0.3$ keV and
$E2\!=\!10$ keV, as long as $E_p(t) \!\gg\! E2\!\geq\!E1$, the energy flux
in an ICS pulse/flare decays like $t^{-2}$ until it is taken over by the
SR afterglow. If $E1\!\ll\! E_p(t)$ but $E2\gsim E_p(t)$ the energy flux
decays like $t^{-4},$ and when $E1\gsim E_p(t)$ the energy flux decays
like $t^{\!-\!4}\, e^{\!-\!E\,t^2/2\, E_p\, t_p^2}$.

\section{The lightcurve of early-time SR flares}

The SR radiation which is emitted from the encounter of a CB with the
wind/ejecta of the progenitor star with a density profile $n(r)\!\propto\!
e^{-a_r/(r\!-\!r_w)}/(r\!-\!r_w)^2$ for $r\!>\!r_w$ and $n(r)\!=\!0$ for 
$r\!<\!r_w$ is 
given
approximately by (DDD2009, DD2008):
\begin{equation}
F_\nu \propto  {e^{-a/t}\, t^{1-\beta} \over t^2+t_{exp}^2}\,
\nu^{-\beta}\,,
\label{SRP}
\end{equation}
where $t\!=\!T\!-\!T_w$ with $T$ being the time after trigger and $T_w$
the time of the CB-wind encounter, $t_{exp}$ is the typical slow-down
time of the fast CB expansion, $\beta\!=\!\Gamma\!-\!1$, and the
exponent describes the decreasing attenuation of the emitted radiation
when the CB penetrates the wind, or more likely, the initial rise
in the wind density due to an exponential cutoff in the wind ejection
if the observed rise in the prompt SRF is achromatic.
(A Gaussian cutoff, $e^{-a^2/t^2}$,  may be required by very 
sharp achromatic SRFs).
Note that for $t^2\!\gg\!t_{exp}^2\,, $
the asymptotic decline of an SRF is a simple power law,
\begin{equation}
F_\nu \propto   t^{-\Gamma}\, \nu^{1-\Gamma}\, ,
\label{SRtail}
\end{equation}
which distinguishes SRFs from ICFs.
This asymptotic decline is insensitive to the exact values
of the start time and the width of the SRF.

Note that as long as $E_p(t)\!\gg\! E$, the temporal decay of ICS
pulses/flares that follows from Eq.~(\ref{ICSlc}), i.e.,
$F_\nu\!\propto\!t^{-2}$, is similar to that of X-ray SRFs that follows
from Eq.~(\ref{SRP}) for $\Gamma_X\!\sim\! 2$, but their spectral indices
differ roughly by one unit ($\Gamma_X\!\sim\! 1$ for ICFs).

Generally, each ICS flare is accompanied/followed by an SR flare
whose lightcurve is given by Eq.~(\ref{SRP}). However 
the SR flares are wider and very often blended 
and the temporal structure is either smotthed or missed because of low 
temporal resolution (see, e.g., Figure~1 in Bartolini et al~2009).

\section{The canonical SR afterglow }

For a constant density ISM and an X-rays well above the cooling 
frequency of the Fermi accelerated electrons in the CBs, the unabsorbed
spectral energy density of their emitted SR is given by 
(DDD2009 and references therein),
\begin{equation}
F_{ISM}[\nu,t] \propto
           \gamma(t)^{3-\beta_X}\, \delta(t)^{3\,\beta_X+1}\,
\nu^{-\beta_X}
\label{Fnu}
\end{equation}
where $\delta\! =\! 1/\gamma\, (1\!-\!\beta\, cos\theta)$ is
its Doppler factor with $\theta$ being the angle between the line of sight
to the CB and its direction of motion. For $\gamma^2 \gg 1$ and $\theta^2
\ll 1$, $\delta \approx 2\, \gamma/(1\!+\!\gamma^2\, \theta^2)$ to an
excellent approximation. In the CB model
the canonical value of the spectral index above
the characteristic frequency has the value $\beta_X\!\approx\!1.1\,.$
For a CB of a baryon number
$N_{_B}$,  a radius $R$ and an initial Lorentz factor $\gamma_0$,
relativistic energy-momentum
conservation yields  the deceleration law of the CB in
an ISM with a constant density $n$ (DDD2009 and references therein):
\begin{equation}
\gamma(t) = {\gamma_0\over [\sqrt{(1+\theta^2\,\gamma_0^2)^2 +t/t_0}
          - \theta^2\,\gamma_0^2]^{1/2}}\,,
\label{goft}
\end{equation}
with $t_0={(1\!+\!z)\, N_{_{\rm B}}/ 8\,c\, n\,\pi\, R^2\,\gamma_0^3}\,.$
As can be seen from Eq.~(\ref{goft}), $\gamma$  and hence $\delta$
change little as long as $t\!\ll\! t_b\!=\![1\!+\gamma_0^2\,
\theta^2]^2\,t_0\,, $ and Eq.~(\ref{Fnu}) yields the {\it `plateau'}
phase of canonical AGs.
For $t\!\gg\!t_b$, $\gamma$  and $\delta$ decrease like $t^{-1/4}\,.$
The transition $\gamma(t)\!\sim\! \gamma_0\! \rightarrow\!\gamma\!\sim\!
\gamma_0\,(t/t_0)^{-1/4}$
around $t_b$
induces a bend, the so called `jet  break',
in the synchrotron AG
from a plateau to an asymptotic power-law
decay,
\begin{equation}
F_{ISM}[\nu,t] \propto t^{-\beta_X-1/2}\,\nu^{-\beta_X}\, .
\label{Asymptotic}
\end{equation}
Thus, the shape of the entire lightcurve of the SR afterglow
after entering the constant density
ISM depends on only three parameters, the  product $\gamma_0\, \theta$,
the deceleration parameter $t_0$ (or the break time $t_b$)
and the spectral index $\beta_X$. The post break decline is
given by the  simple power-law (Eq. \ref{Asymptotic}) independent of
the values of $\gamma_0\, \theta$ and $t_b$.
In cases where $t_b$ is earlier
than the beginning of the XRT observations or is hidden under
the prompt emission, the entire observed lightcurve of the AG
has this asymptotic power-law form (DDD2008a).

For a wind density profile, $n\!\propto\! 1/r^2$ beyond $r\!=\!r_w$, the 
asymptotic
decline is given by
\begin{equation}
F_{W}[\nu,t] \propto t^{-\beta_X-1}\,\nu^{-\beta_X}\, ,
\label{Asymptoticrm2}
\end{equation}
where $t$ is the time after the onset of the $n\!\propto\! 1/r^2$
density. This relation  describes well the asymptotic decay
of late-time SRFs (DDD2003; DDD2009).

\section{GRB090812} 

\noindent
{\bf Observations:} 
The Swift Burst Alert Telescope (BAT) triggered and located GRB 090812 on
August 12, 2009 at 06:02:08 UT (Stamatikos et al.~2009). 
The BAT light curve (energy flux in the 
15-350 keV range) showed three peaks
around 5, 27 and 55 sec, respectively, after trigger 
(Baumgartner et al.~2009).
The total duration
of the burst was approximately 70 sec. 
A joint spectral analysis of the Konus-Wind and Swift/BAT
time integrated spectrum in the 23-1400 keV energy band 
was well fit  
with a power-law with exponential cutoff 
with a power-law index -1.03 +/- 0.07 and peak energy 
$E_p$=572(-159, +251) keV (Pal'shin et al.~2009).
The Swift
X-ray telescope (XRT) began observing the field at 06:03:25.7 UT, 76.8
seconds after the BAT trigger. The X-ray lightcurve in the 0.3-10 keV
band that was inferred from the Swift XRT observations is shown in 
Fig.~1a. It
was reproduced from the Swift/XRT GRB lightcurve repository ( 
{\it http://www.swift.ac.uk/xrt$\_$curves/}, Evans et al.~2007,2009).
It shows the X-ray tail of the last pulse detected by the BAT around
55 sec followed by two prominent X-ray flares which are taken over around
600s by a smooth $\!\sim\!t^{-2}$ decline until a gap in the data between
1.15 ks and 11.5 ks after which the light curve decay like
a post-break power-law $\!\sim\!t^{-1.4}.$ Inspection of the hardness
ratio measured by the XRT and reported in the Swift/XRT GRB lightcurve
repository reveals the typical fast decay of  flares ending the 
prompt  emission that is accompanied by a rapid spectral
softening. The fast falling hardness ratio increases back to a 
constant
value around 600 sec and remains so during the rest of the
afterglow observations with the XRT. The constant photon spectral index 
during this phase is (Swift/XRT GRB lightcurve Repository): 
$\Gamma_X$=1.914 (+0.138, -0.089).

The optical AG of GRB090812 was first detected 24 s after the BAT trigger 
by the RAPTOR telescope (Wren et al.~2009). A redshift $z$=2.542 was 
inferred for GRB 090812 from early-time observations using the FORS2 at 
the VLT (de Ugarte Postigo et al.~2009). Its optical AG was also detected 
and followed-up by the Swift UVO telescope (Stamatikos et al.~2009; Schady 
et al.~2009), the automated Palomar 60 inch telescope (Cenko et al.~2009), 
the 2 m Liverpool automated Telescope (Smith et al.~2009), the 2.2 m 
ESO/MPI telescope (Updike et al.~2009), and the Faulkes Telescope South 
(Cano et al.~2009). The R-band light curve from these GCN reports is shown 
in Fig~2. It reveals a brightening AG until $\!\sim\!$70s after the 
trigger that turned into a power-law decay. Such behaviour
is typical of the prompt optical emission detected 
with robotic telescopes, by now in many bright GRBs 
(see, e.g., Fig.~3) such as
090123 (Akerlof et al.~1999), 030418 (Rykoff et al.~2004), 050820A
(Cenko et al~2006), 060418 (Molinari et al.~2007), 
060605 (Ferrero et al.~2009), 060607A (Molinari et al.~2007; 
Ziaeepour et al.~2008; Covino et al.~2008a; Nysewander et al.~2009),
061007 (Mundell et al.~2007),  071010A (Covino et al.~2008b) 081203A, 
(Kuin et al.~2009), 090102 (Klotz et al.~2009a,b; Covino et al.~2009),  and 
090618 (Li et al.~2009).  

\noindent
{\bf CB model interpretation of the XRT lightcurve:} The spectral 
evolution of the X-ray tail of 
the last flare detected by the Swift BAT around 55 sec and of the 
following two X-ray flares detected by the Swift XRT are that expected in 
the CB model for ICFs (DDD2009, DD2009b).
Consequently, we have reproduced the early-time XRT 
lightcurve by a sum of an X-ray tail and two prominent ICFs 
(Eq.~(\ref{ICSlc})) which is taken over by the tails of the SRFs 
(Eq.~(\ref{SRP})) that are associated with these ICFs.  
The late time X-ray 
afterglow was reproduced by a CB model post-break SR afterglow, as given 
by Eqs.~(\ref{Fnu}) and (\ref{goft}). 
The values of the parameters used in the CB model description 
of the complete X-ray lightcurve are 
listed 
in Table~\ref{t1}. Because of the approximations, the possibility 
of local minima in the $\chi^2$ search and 
degeneracy of parameters,  
the best fit values of the parameters, probably 
are effective (approximate) values  
and are not necessarily their exact physical values. 
For instance, the tail of the first ICF is not sensitive to 
its beginning time and its width. The joint tail of the SRFs depends only 
on $\beta_X$. The spectral index 
parameter $p$ in 
the description of the AG was constrained to satisfy the CB model closure 
relation for GRB090812: $p/2\!=\!\beta_x\!=\!\Gamma_X$-1=0.914 (+0.138, 
-0.089). However, because of the gap in the XRT data between 1.15 ks and 
11.5 ks and the large uncertainty in the value of $\beta_X$ inferred from 
the data, the values of $p$, $\gamma_0\theta$ are not well determined by 
the fit. Thus, in the CB-model description of the late-time X-ray AG we 
adopted the central value reported in the Swift/XRT GRB lightcurve 
repository, $\beta_X\!=\!0.914$, which yields a post-break behaviour, 
$F_\nu\!\propto\! t^{\!-\!1.43}$ and the best fit values 
$\gamma_0\,\theta$=1.34 and $t_b$=598 sec. The CB-model description of the 
complete XRT lightcurve is shown in Fig.~1a. An enlarged view of the early 
time behaviour is shown in Fig.~1b. This CB model description yields 
$\chi^2/dof$=481/421=1.18.

\noindent
{\bf CB model interpretation of the optical lightcurve:} The data on the 
optical lightcurve of GRB090812 that was reported by different groups in 
GCN reports are preleminary, sparse, was neither cross calibrated nor 
corrected for extinction along the whole line of sight. Despite that, the 
data show roughly the behaviour predicted by the CB model, which is 
demonstrated in Fig.~2. The CB model description assumes a prompt emission 
SRF from the encounter of the CBs with the wind/ejecta blown by the 
progenitor star long before the GRB, which begins in the observer frame 
towards the end of the first $\gamma$-ray flare detected by the Swift BAT, 
and turns into a power-law decline (Eq.~\ref{SRP}) that is taken over by 
the achromatic SR afterglow ($\beta_O\!=\!\beta_X$, 
$\gamma_0\,\theta$=1.34 and $t_b$=598 sec obtained from the CB-model fit 
to the 
X-ray AG). The broad SRF is probably dominated by a sum of 3 unresolved 
SRFs associated with the 3 prominent BAT flares. and the 2 prominent XRT 
flares which are blended together into a broad SRF. The effective 
parameters which were used in the CB-model description of the prompt 
emission SRF are $t_0\!\sim\!10$ sec, $t_{exp}\!\sim\!66$ sec, 
$a\!\sim\!21.3$ sec and $\beta_O\!\sim\!0.55$.
Only in very bright GRBs, such as 080319B 
the prompt optical emission is resolved into separate SRFs 
(Racusin et al.~2008),
associated with the prompt $\gamma$-ray flares, as shown in Fig.~2b
borrowed from DD2008.

{\bf Conclusion:} The lightcurves and spectral evolution of the X-ray
and optical flares in GRB090812 are well reproduced by the CB model.

 {}
\begin{deluxetable}{llllc}
\tablewidth{0pt}
\tablecaption{The parameters of the ICFs  used in the CB model
description of X-ray  lightcurves of Swift GRBs.}
\tablehead{\colhead{flare}& \colhead{$t_i$ [s]} & \colhead{$\Delta t$} &
\colhead{$E_p(0)$ [keV]} & \colhead{A [erg/cm$^2$~s$^{-1}$}] }
\startdata
ICF1 & 40.3 & 25.6 & 9.76& $0.26\time 10^{-7}$ \\

ICF2 & 90.4 & 54.3 & 0.45& $0.23\time 10^{-6}$ \\

ICF3 & 236.8 & 35.4 & 4.30& $0.80\time 10^{-8}$ \\

\hline                                          
 SRF  & $t_i$ [s] & $t_{exp}$  & a [s] &       \\             
 ---  &  125.4    & 165.4      & 2.0   &       \\    
\hline 
   AG &  $\gamma_0 \theta$ & $t_0$ [s]   &  $\beta_X$ &    \\    
      &   1.095            &  182      &   0.98  &        \\  
\enddata
\label{t1}
\end{deluxetable}

\begin{figure}[]
\centering
\vspace{-1cm}
\vbox{
\epsfig{file=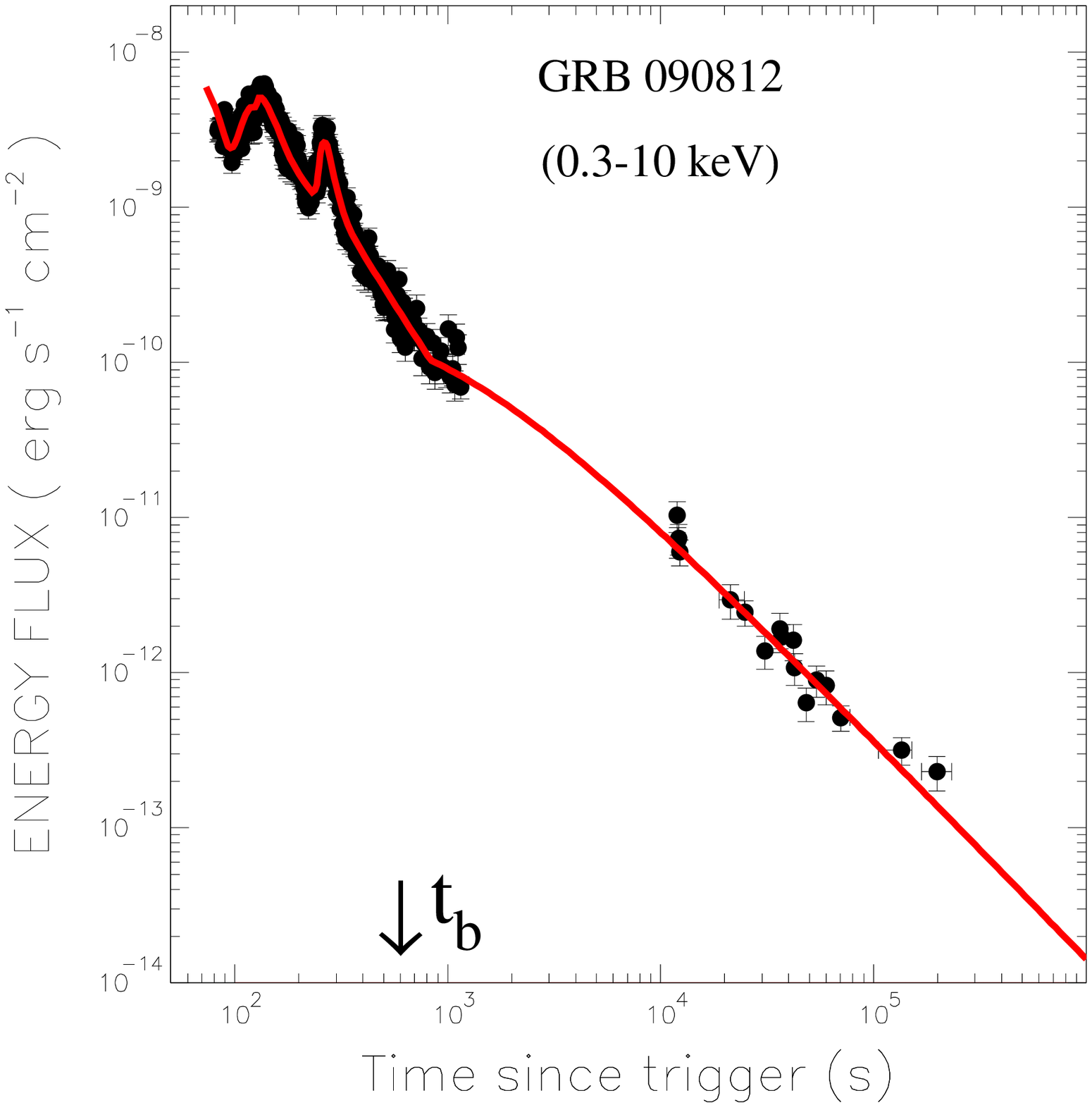,width=14.0cm,height=11.0cm}
\epsfig{file=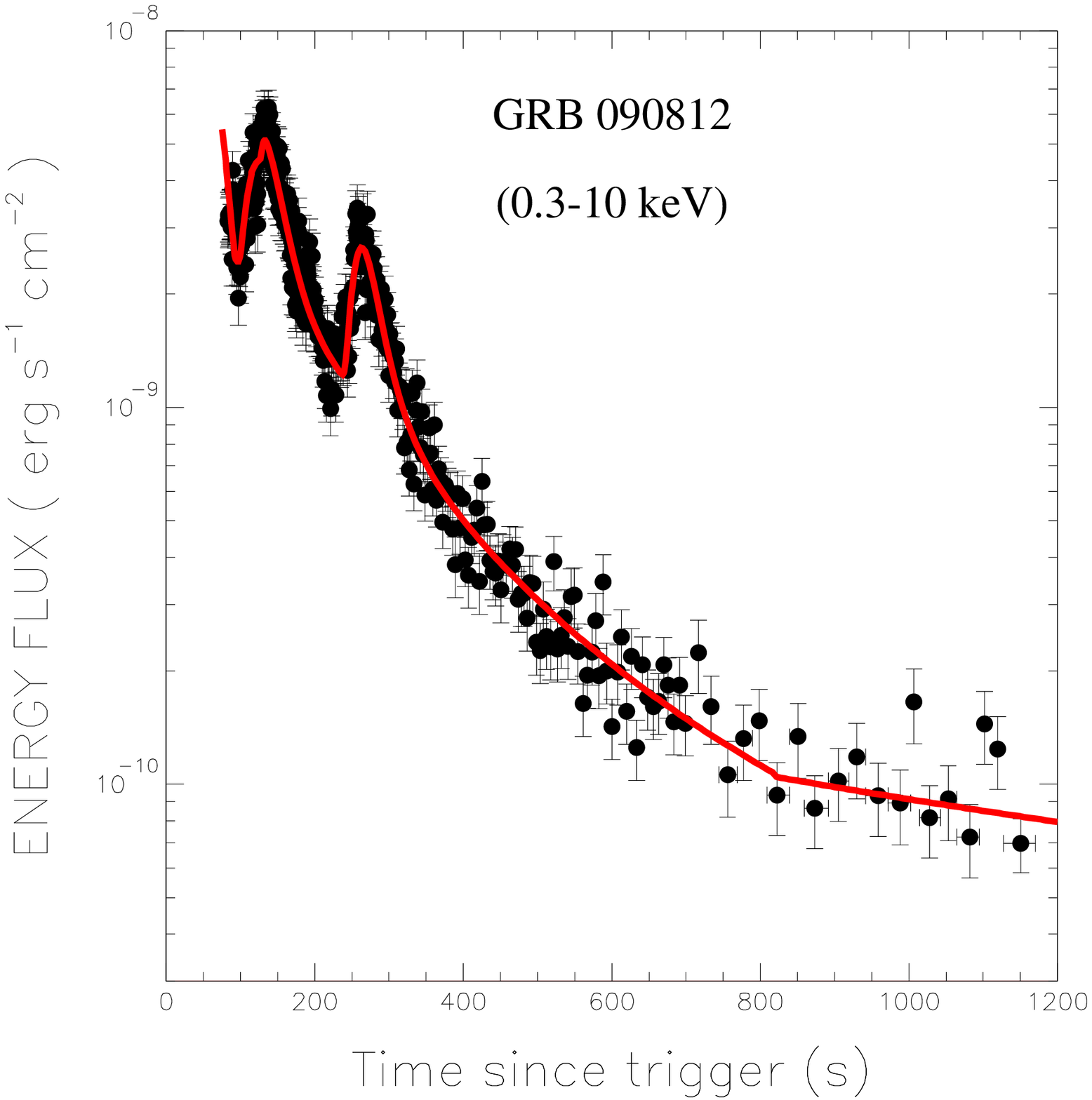,width=14cm,height=11.0cm}
}
\caption{{\bf Top (a):}
Comparison between the 0.3-10 KeV X-ray lightcurve
of GRB090812
measured with the Swift XRT and reported in the Swift/XRT lightcurve
repository {\it http://www.swift.ac.uk/xrt$_{-}$curves/}  (Evans et
al.~2009) and its CB model description 
as detailed in the text.
{\bf Bottom (b):}  Enlarged view of the comparison in part {\bf(a)}
for the early time flares.
}
\label{fig1}
\end{figure}

\newpage
\newpage
\begin{figure}[]
\centering
\vspace{-1cm}
\vbox{
\epsfig{file=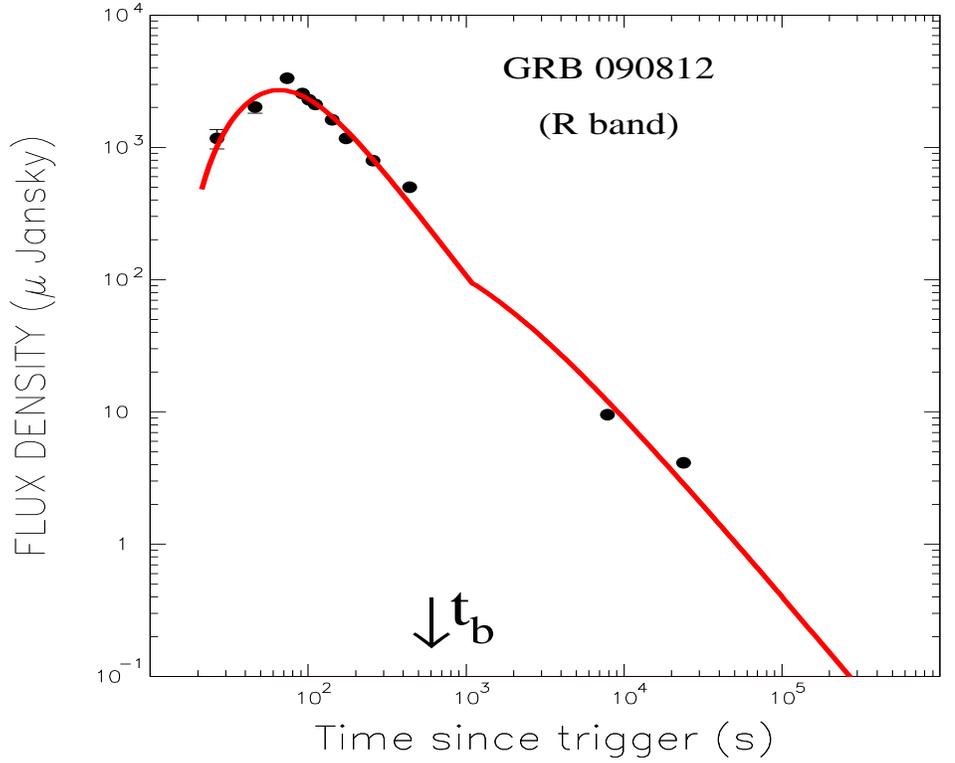,width=14.0cm,height=11.0cm}
\epsfig{file=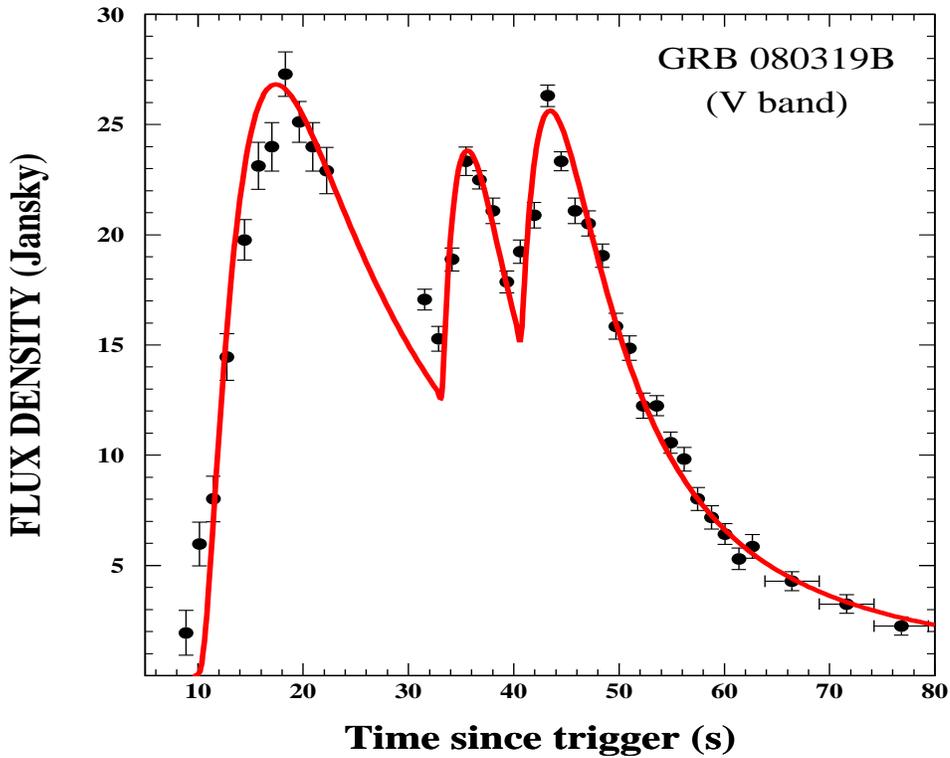,width=14.0cm,height=11.0cm}
}

\caption{{\bf Top (a):} Comparison between the $R$-band lightcurve of the 
prompt optical emission in the bright GRB 090812 and its CB model 
description as detailed in the text. {\bf Bottom (b):} Comparison between 
the V-band lightcurve of the prompt optical emission in GRB080319B, the 
brightest observed GRB so far, and its CB model description as detailed in 
DD2008. The smooth lightcurve of GRB090812, in contrast to that of 
GRB080319B, may result from low temporal resolution due to low 
statistics 
and/or due to blending of the individual optical flares 
following/associated with its prompt $\gamma$-ray flares.
}
\label{fig2}
\end{figure}

\newpage
\begin{figure}[]
\centering
\vspace{-1cm}
\vbox{
\hbox{
\epsfig{file=R090812.eps,width=8.0cm,height=6.0cm}
\epsfig{file=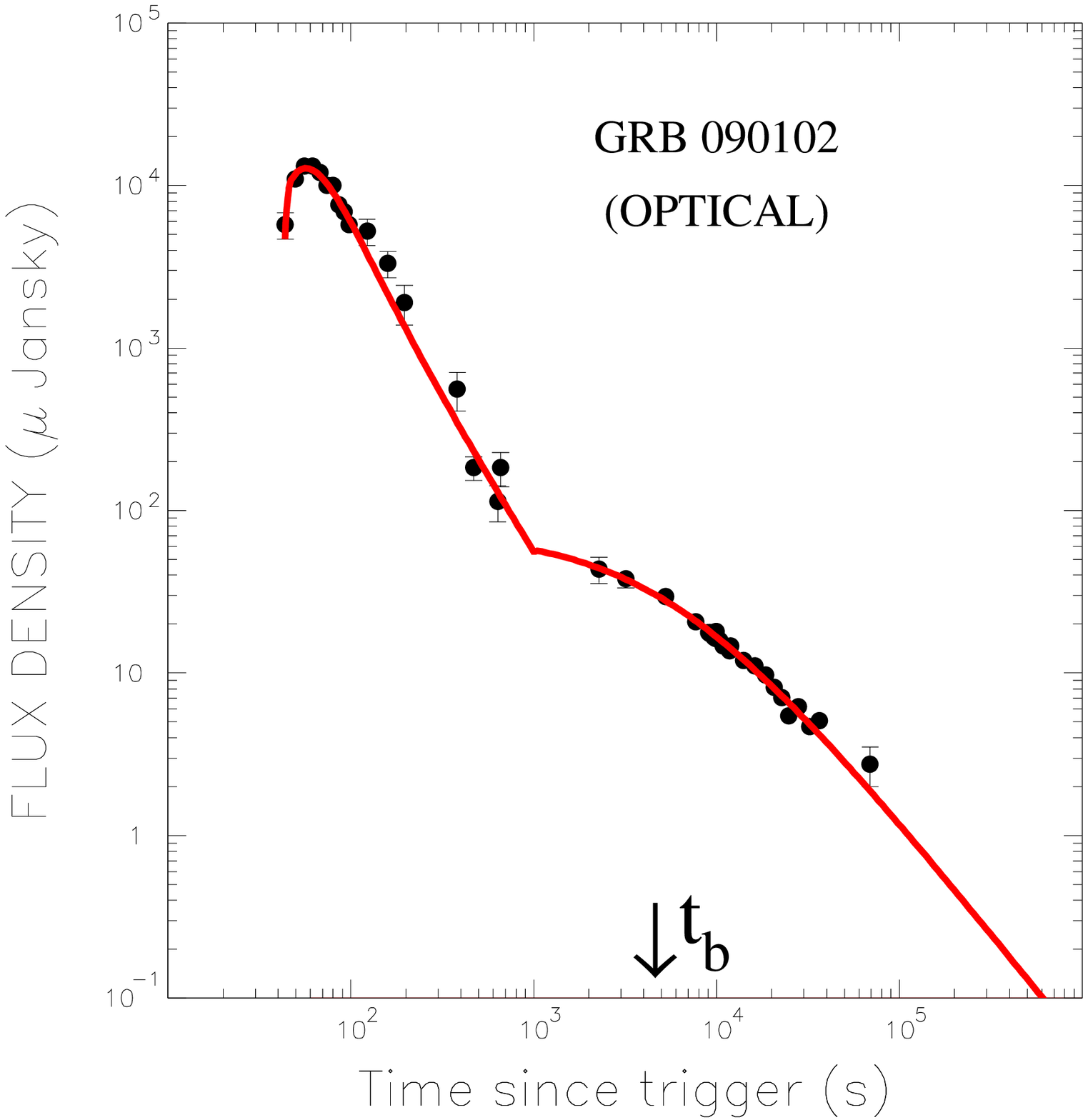,width=8.0cm,height=6.0cm}
}}
\vbox{
\hbox{
\epsfig{file=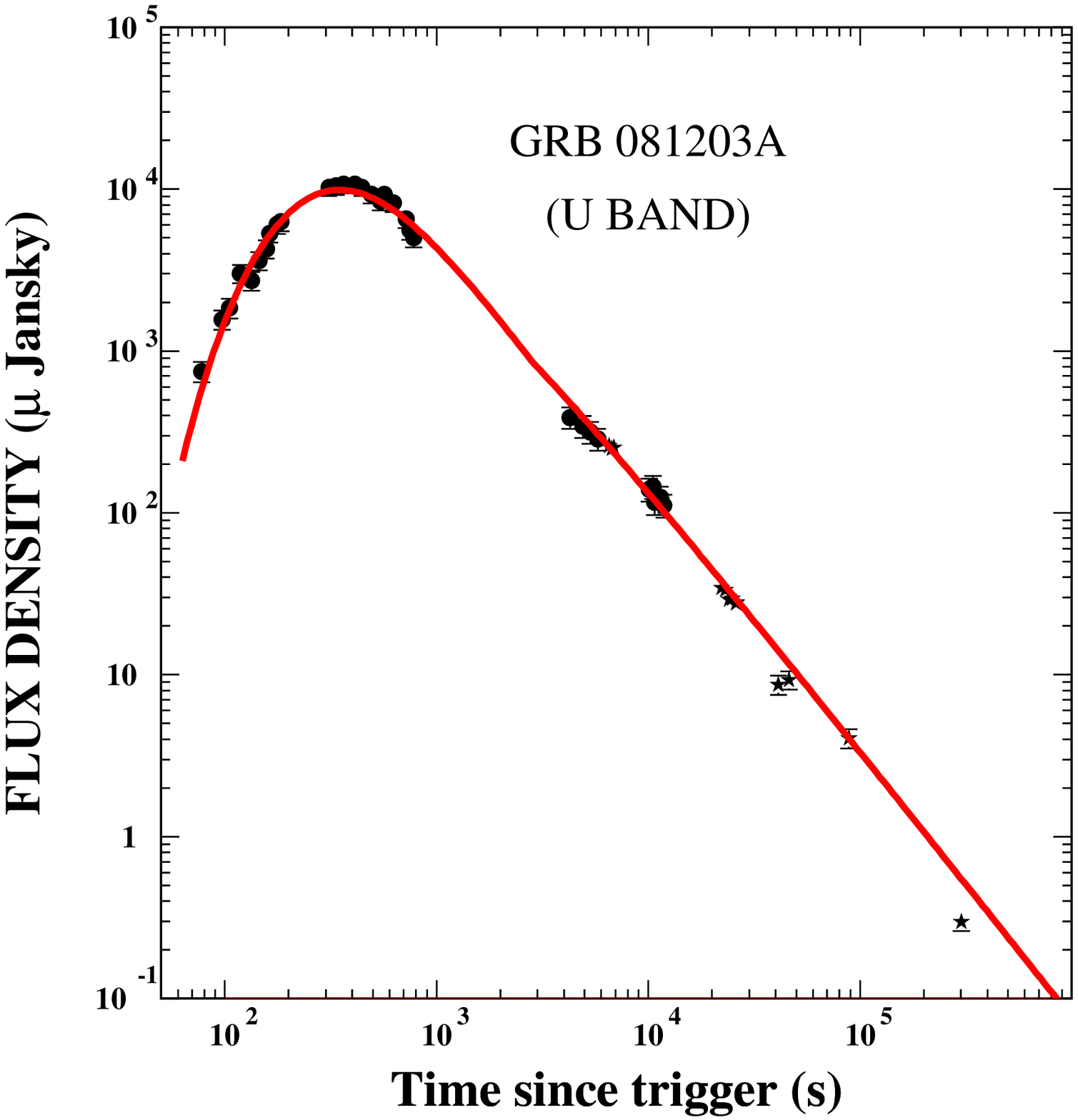,width=8.0cm,height=6.0cm}
\epsfig{file=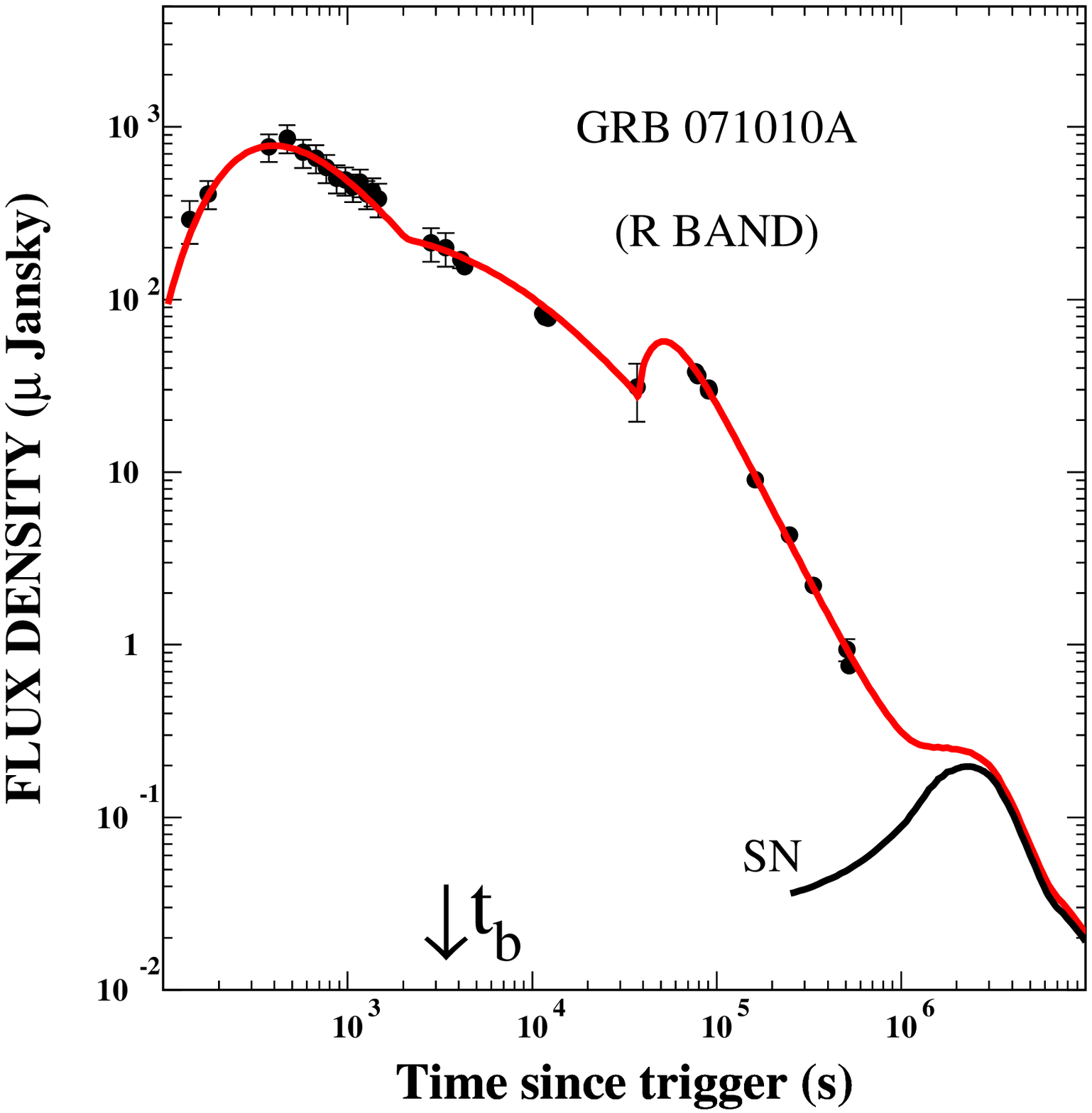,width=8.0cm,height=6.0cm}
}}
\vbox{
\hbox{
\epsfig{file=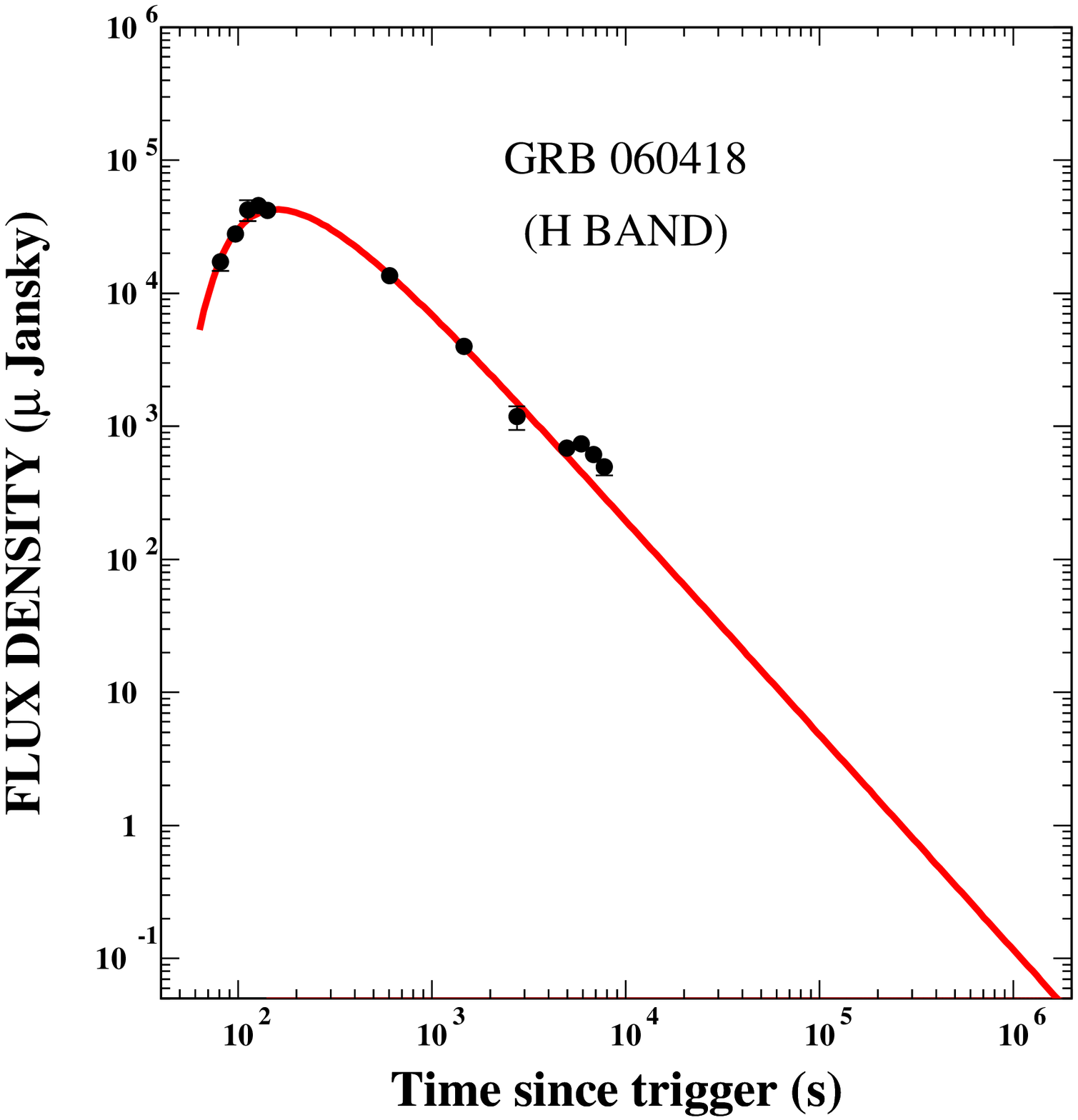,width=8.0cm,height=6cm }
\epsfig{file=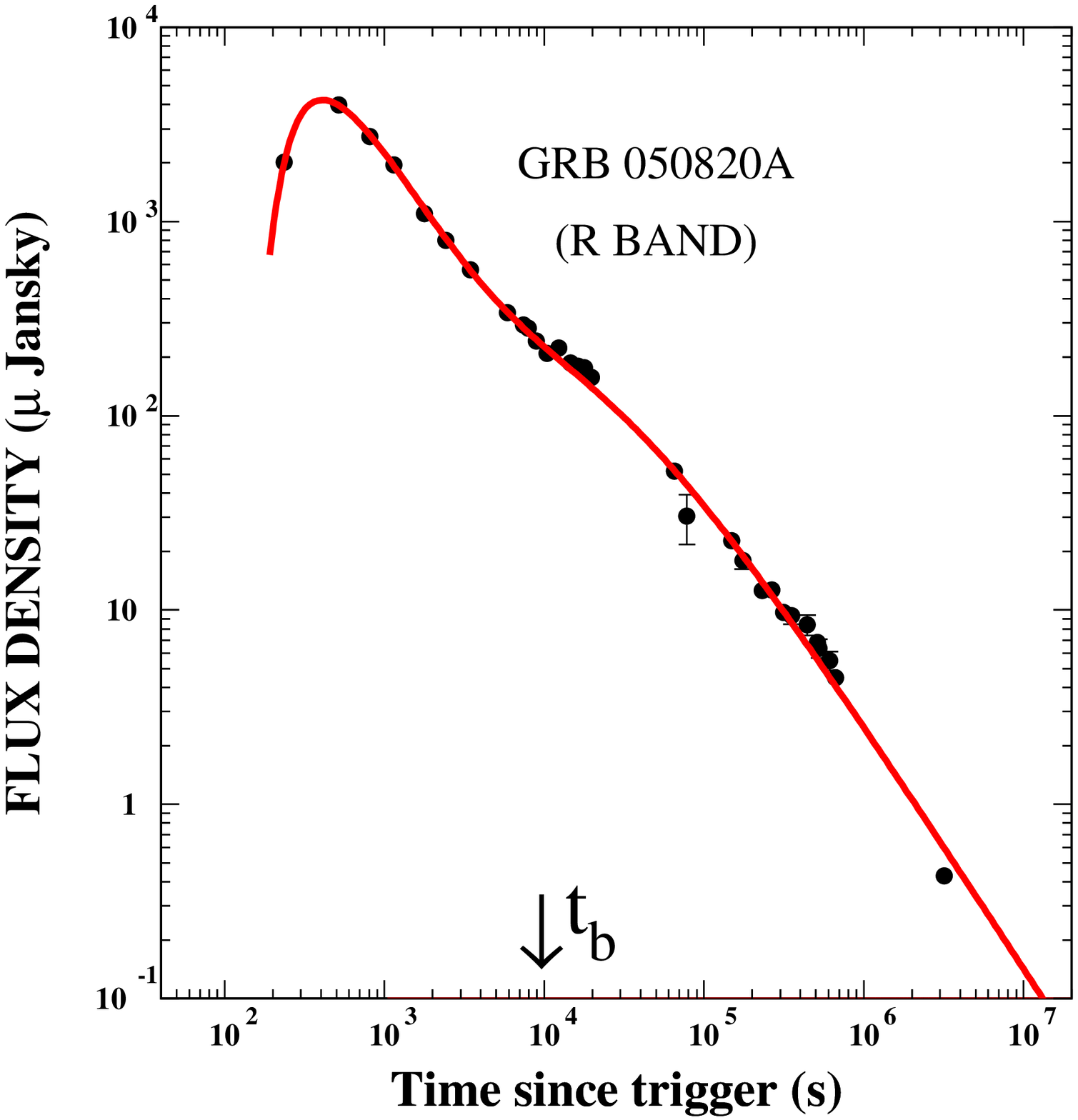,width=8.0cm,height=6cm}
}}
\caption{The R-band lightcurve of GRB090812 
and a few similar optical lightcurves of other GRBs which were detected by 
automated 
telescopes 
during the prompt emission phase of these GRBs. Also shown is their 
CB-model description (with exponential cutoffs of wind ejections).
{\bf Top left (a):}     GRB 090812.   See text for details.
{\bf Top right (b):}    GRB 090102.   Figure borrowed from DD2009a. 
{\bf Middle left (c):}  GRB 081203A.  Figure borrowed from DD2009a.
{\bf Middle right (d):} GRB 071010A.  Figure borrowed from DDD2009.
{\bf Bottom left (e):}  GRB 060418.   Figure borrowed from DDD2009.
{\bf Bottom right (f):} GRB 050820A.  Figure borrowed from DDD2009.
}
\label{fig3}
\end{figure}

\end{document}